\documentclass{iau} 
\usepackage{graphicx}
\usepackage{natbib}

\title[Deep Machine Learning in Cosmology]
{Deep Machine Learning  in Cosmology: Evolution or Revolution?}

{

\author[Ofer Lahav]
{ Ofer Lahav$^1$
}

\affiliation{
  $^2$Department of Physics and Astronomy, University College London, London WC1E 6BT, UK\\
 }

\pubyear{2022}
\volume{368}  
\setcounter{page}{1}
\jname{Machine Learning in Astronomy: Possibilities and Pitfalls}
\editors{A. Mahabal, C. Fluke \& J. McIver, eds.}
\begin{document}

\maketitle

\begin{abstract}
Could Machine Learning (ML) make fundamental discoveries and tackle unsolved problems in Cosmology? 
Detailed observations of the present contents of the universe are consistent with the Cosmological Constant $\Lambda$ \& Cold Dark Matter model, subject to some unresolved inconsistencies (‘tensions') among observations of the Hubble Constant and the clumpiness factor. To understand these issues further, large surveys of billions of galaxies and other probes require new statistical approaches. In recent years the power of  ML, and in particular `Deep Learning',  has been demonstrated  for object classification, photometric redshifts, anomaly detection, enhanced simulations,  and inference of cosmological parameters. 
It is argued that  the more traditional 'shallow learning' (i.e. with pre-processing feature extraction) is actually quite deep, as it brings in human knowledge, while 'deep learning'  might  be perceived as a black box, unless supplemented by explainability tools. 
The `killer applications' of ML for Cosmology are still to come.
New ways to train the next generation of scientists for the Data Intensive Science challenges ahead are also discussed. Finally, the chatbot ChatGPT  is challenged to address the question posed in this article's title.
\footnote{This article is based on presentations of the Royal Astronomical Society George Darwin Lecture (2020), a plenary talk at the IAU Symposium 368 held in South Korea (2022) and a talk at the workshop `Unsolved Problems in Astrophysics and Cosmology' at the Hebrew University (2022) .} 
  \end{abstract}

\keywords{methods: data analysis -- methods: statistical -- techniques: galaxies}

\firstsection 
\section{Introduction}

The exponential growth of Astronomical data is remarkable: for example, from samples of thousands and tens of thousands of galaxies in the 1980s (CfA, IRAS) to millions of galaxies in the 1990s and early 2000s (2dF, SDSS, BOSS), to 300 million galaxies at present (DES), and to billions of galaxies in new surveys (Rubin-LSST, Euclid) which will start in ~2023-24. The data acquisition rate, number of galaxies, cost and number of researchers are given in Figure 1 below for some of the surveys.

\begin{figure}
\begin{center}
  \begin{minipage}{140mm}
  \centering
  \raisebox{-0.5\height}{\includegraphics[width=150mm, angle=0] {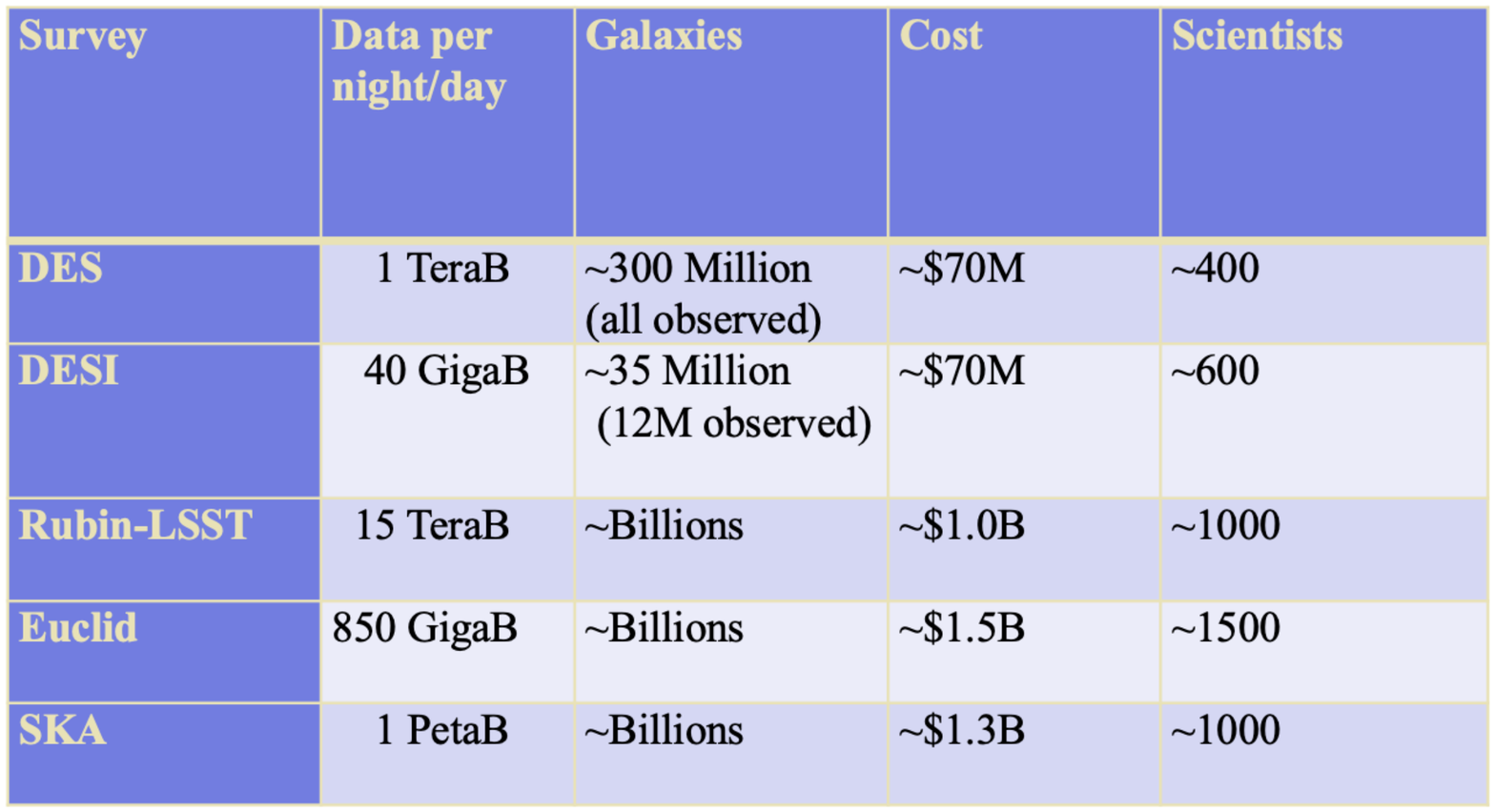}}
 
  \end{minipage}
  \caption{Examples of galaxy surveys: data rates, number of galaxies, financial resources and human capital. DES observations are complete, DESI is collecting data, and Rubin-LSST, Euclid and SKA will be operating coming years.}
   \label{fig1}
\end{center}
\end{figure}
The other revolution, across the whole of science and technology, is in the developments of Artificial Intelligence (AI) and Machine Learning (ML).  Terminology-wise, ML is a subset of AI, and Deep Learning (DL) is a new approach within  ML (see Figure 2). 
\begin{figure}
\begin{center}
  \begin{minipage}{140mm}
  \centering
\raisebox{-0.3\height}{\includegraphics[width=90mm]{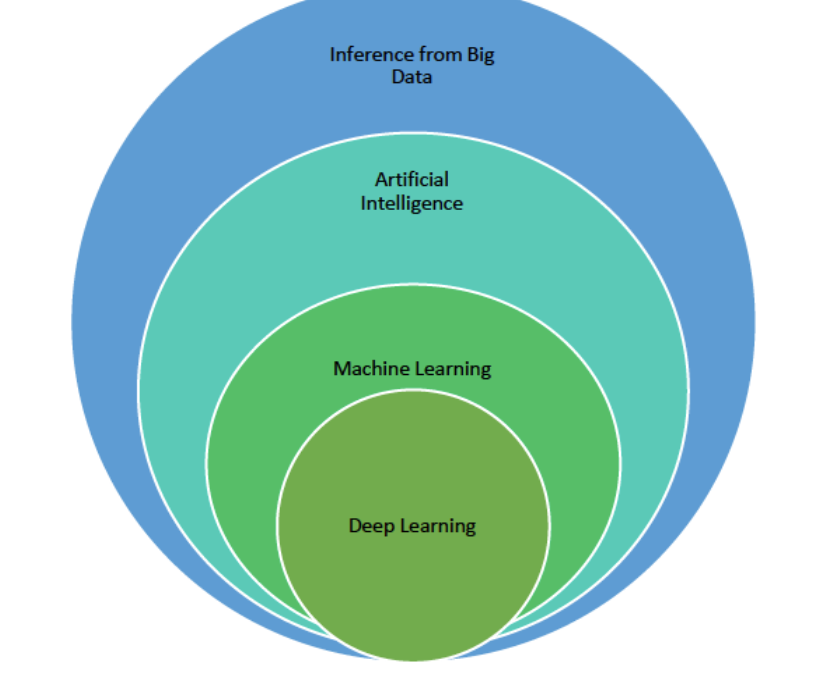}}
  \hspace*{.4in}
    \vspace*{+3mm}
  \raisebox{-0.7\height}{\includegraphics[width=100mm]{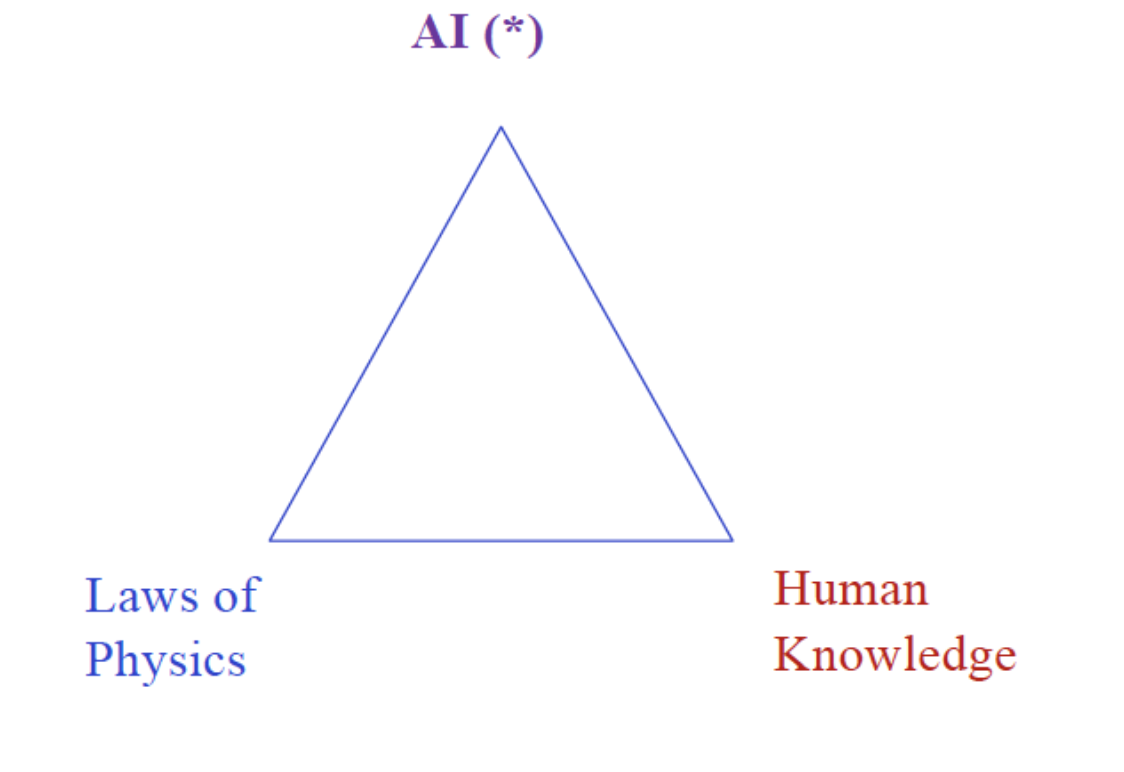}}
  \end{minipage}
  \vspace*{+3mm}
  \caption{Top: The relationship between sub-fields: Statistical Inference, Artificial Intelligence, Machine Learning and Deep Learning, Bottom: the interplay between (1)  'objective' laws of Physics (2) current Human knowledge and (3)  Augmented Intelligence  that could  update human knowledge  based on new data.}
   \label{fig1}
\end{center}
\end{figure}
AI \&ML in Astronomy is a very busy field now, with many papers published weekly, beyond what can be shown in a short presentation. I am taking the liberty to present here work over the years with my PhD students and Post-docs, and to use these as examples for a general discussion of AI \& ML in Astronomy and beyond. 
While the application of ML in Astronomy is very common these days,  there are open issues to what extent they are  `black boxes’, the connection to traditional statistics 
and how to implement these methods.
For recent comprehensive reviews see  Baron (2019) on applications of ML to Astronomy, Carleo et al. (2019) 
on ML  across the Physical Sciences, and  
 Huertas-Company \& Lanusse (2022), on DL for galaxy surveys, with applications for object classification, galaxy properties, discovery, emulation of simulations and inference.  

The outline of this article is as follows. 
The status of the $\Lambda$ \& Cold Dark Matter ($\Lambda$CDM) model is reviewed in the context of present and future galaxy surveys, that will chart billions of galaxies. Examples of AI \& ML in Cosmology and Astrophysics are presented: (i) Object classification;  (ii) Photometric redshifts;  and  (iii) Dark matter mapping
and (iv) Simulation-based inference of the masses of the Milky Way and Andromeda and of cosmological parameters.
The training of the next generation of astronomers as data scientists and the outlook for AI\&ML in Astronomy is reviewed, with the example of UCL's Centre for Doctoral Training (CDT) in Data Intensive Science (DIS).

\begin{figure}
\begin{center}
  \begin{minipage}{140mm}
  \centering
 \raisebox{-0.5\height}{\includegraphics[width=100mm, angle=0] {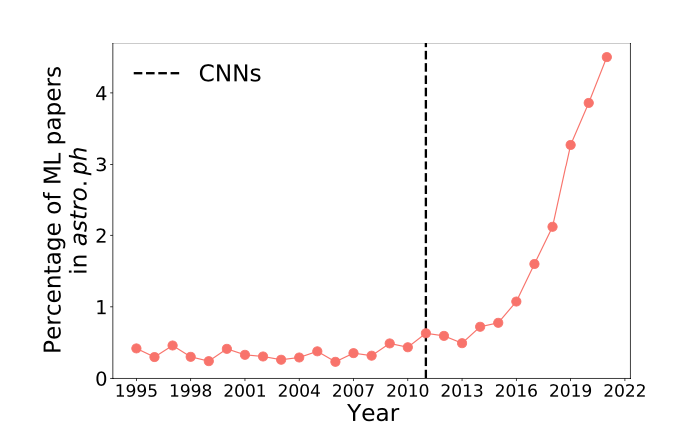}}
  \hspace*{.2in}
  \raisebox{-0.5\height}{\includegraphics[width=100mm]{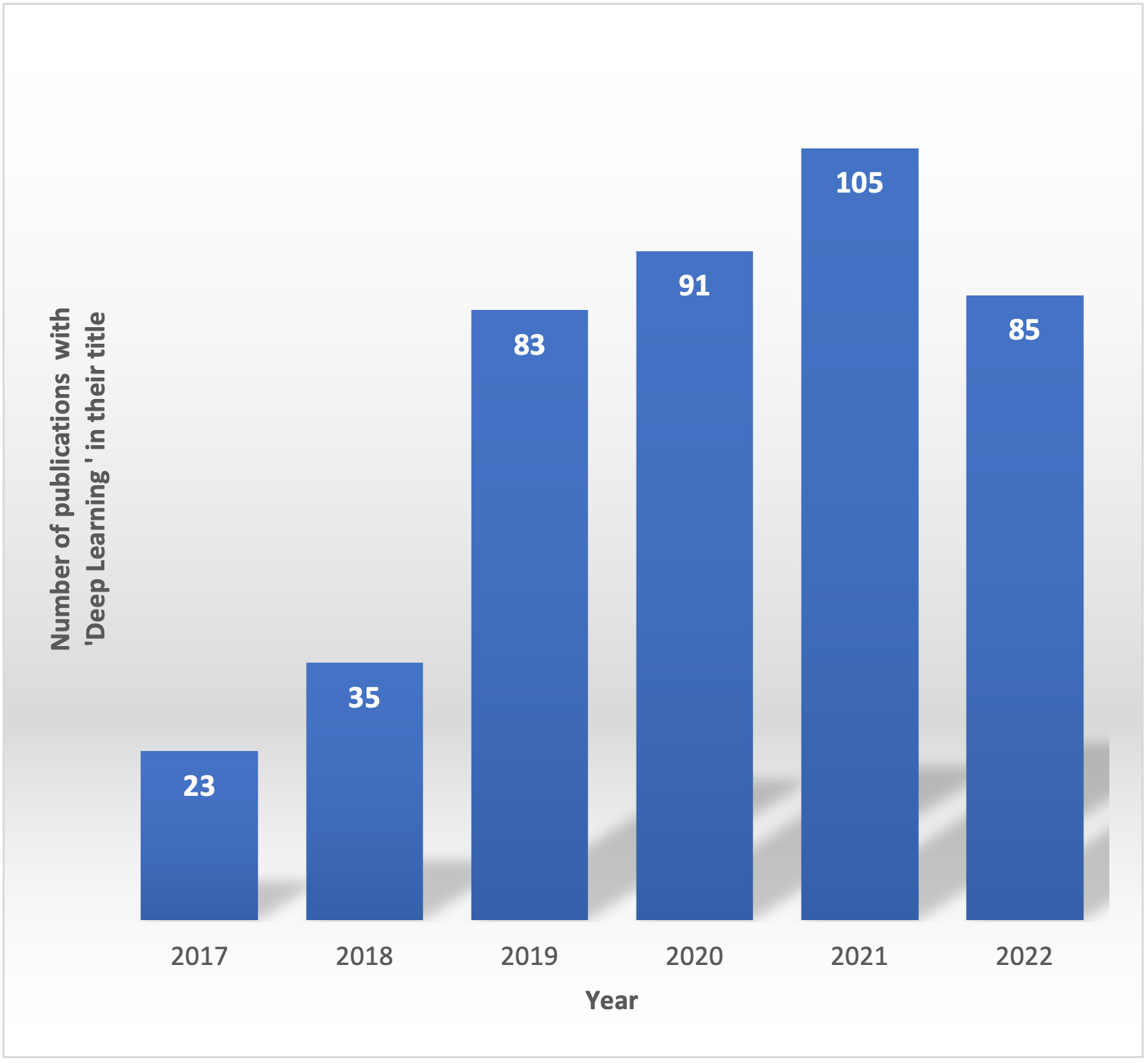}}
  \end{minipage}
  \vspace*{+3 mm}
  \caption{Top:The percentage of papers on astro-ph (arXiv.org) that report on at least one  type of ML technique in the title or abstract.  Some papers on ML for Astronomy appeared already prior to 1995 (see the text).
  The vertical line indicates the popularisation of Convolutional Neural Networks. 
  Credit: D. Piras, PhD thesis (CDT-DIS/UCL). Bottom: The papers on astro-ph arXiv  per year with `Deep Learning' in their title.}
   \label{fig2}
\end{center}
\end{figure}


\section{Background: the status of the cosmological model }




Recent years have witnessed success after success of Einstein’s General Relativity: from the measurement of a Cosmological Constant $\Lambda$ (or its variants) to the detection of Gravitational Waves (GW) moving at the speed of light (LIGO collaboration 2017a). In more detail, recent cosmological measurements strongly favour a ‘concordance’ model in which the present-time universe is flat and contains approximately 5\% baryons, 25\% cold dark matter and 70\% dark energy, with a small contribution from massive neutrinos (e.g. Planck collaboration 2018; Dark Energy Survey collaboration 2021; Lahav \& Liddle 2022 for a review). The modern observational evidence for dark energy is attributed to Supernovae Ia surveys in the late 1990s (Perlmutter et al. 1999; Riess et al. 1998), with subsequent support from other observational probes, although there was already growing evidence for it in the previous decade. The concept of dark energy is a variant on Einstein's cosmological constant, $\Lambda$, and the proposition for a $\Lambda$-like linear force can be traced back to Newton's Principia  (e.g. Calder and Lahav 2008). This $\Lambda$CDM paradigm and its extensions pose fundamental questions about the origins of the universe.
Commonly, dark energy is quantified by an equation of state parameter $w$, which is the ratio of pressure to density, $w=P/\rho$. The case $w=-1$ corresponds exactly to Einstein's Cosmological Constant in General Relativity. But $w$ may vary with cosmic epoch, e.g. in the case of scalar fields.

Essentially $w$ affects both the geometry of the universe and the growth rate of structures. These effects can be observed via a range of cosmological probes, including the Cosmic Microwave Background (CMB), galaxy clustering, clusters of galaxies, weak gravitational lensing and Supernovae Ia. For example, Supernovae Ia act as `standard candles', while Baryonic Acoustic Oscillations (BAO) serve as `standard rulers’. 
The current cosmological model, $\Lambda$CDM, has been spectacularly successful in fitting most of the observations. However, there is tension, at about 4$\sigma$, between the Hubble Constant $H_0 = 67.4 \pm  0.5$ km/sec/Mpc from CMB measurements (Planck collaboration 2018) and that measured locally $H_0 = 73.2 \pm 1.3$ km/sec/Mpc (Riess et al. 2021), that could force us to reconsider the standard $\Lambda$CDM model (for review see Shah et al. 2021). Unknown systematic errors might be the cause of this tension, and so we are in need of an independent probe of $H_0$ (e.g. gravitational wave `bright’ and `dark’ sirens) to confirm whether our model of the universe is in trouble. There is also tension, of about 2$\sigma$ in the clumpiness parameter $S_8$, in the sense that the mass fluctuations measured by weak lensing in three different surveys seem a bit `smoother’ than what is measured by the Planck CMB.	
Could Machine Learning help to tackle these problems in Cosmology? Some challenges and examples of ML applications are discussed below.

As an example of one of the surveys we highlight now the Dark Energy Survey (DES), in which I have been involved since 2004. DES completed in 2019 six observing seasons, over 758 nights. DES has used a wide-field camera on the 4m Blanco Telescope in Chile to image 5000 sq
deg of the sky in five filters. The survey has generated a catalogue of 300 million galaxies with photometric redshifts and 100 million stars. In addition, a time-domain survey search over 27 sq deg has yielded a sample of thousands of Type Ia supernovae and other transients.  The DES collaboration (2021) has measured $w = -1.03 \pm  0.03$ (68\% CL)  (using DES weak lensing \& galaxy clustering combined with Planck and other probes, consistent with Einstein’s Cosmological Constant. DES has also yielded unexpected science, including the discovery of distant objects in the Solar System, dwarf galaxies of the Milky Way, high-redshift Quasars and the detection of an optical flash from the LIGO Gravitational Wave Binary Neutron Star event GW170817. The story of DES, from an idea in 2003 through construction and observations to the first results in 2019 is described in the DES book (eds. Lahav et al. 2020). 

The science of the next generation of surveys, ‘Stage IV’ (Rubin-LSST, DESI, Euclid, Roman-WFIRST) is even more ambitious. An intriguing question is  ``when to stop?" to measure cosmological  parameters such as $w$ (e.g. Lahav \& Silk 2021). The short answer is that it depends what are the alternative theories. There is of course merit in increasing accuracy and precision of the measurements, which may reveal unexpected results. 
Detecting deviations from $\Lambda$CDM and measuring the neutrino mass would potentially be  landmark measurements for  the whole of Physics.


\bigskip
\bigskip

\section{Deep vs. Shallow Learning}

\begin{figure}
\begin{center}
  \begin{minipage}{140mm}
  \centering
 \raisebox{-0.5\height}{\includegraphics[width=140mm, angle=0] {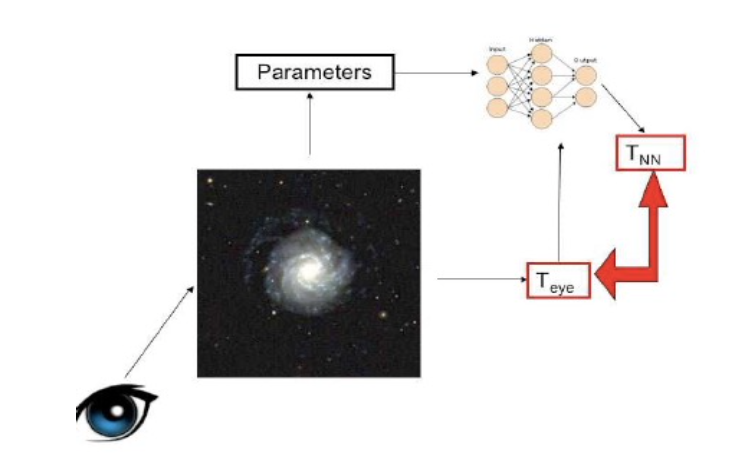}}
  \hspace*{.2in}
  \end{minipage}
  \vspace*{+3 mm}
  \caption{An illustration of supervised learning with Artificial  Neural Networks for galaxy classification. The classification of a training set is done by visual inspection of the entire image, 
  while the training and testing only use a dozen or so  features extracted by pre-processing from the galaxy images. Graphics Credit: M. Banerji et al. (2010). }
   \label{fig2}
\end{center}
\end{figure}

\begin{figure}
\begin{center}
  \begin{minipage}{140mm}
  \centering
  \raisebox{-0.5\height}{\includegraphics[width=140mm, angle=0] {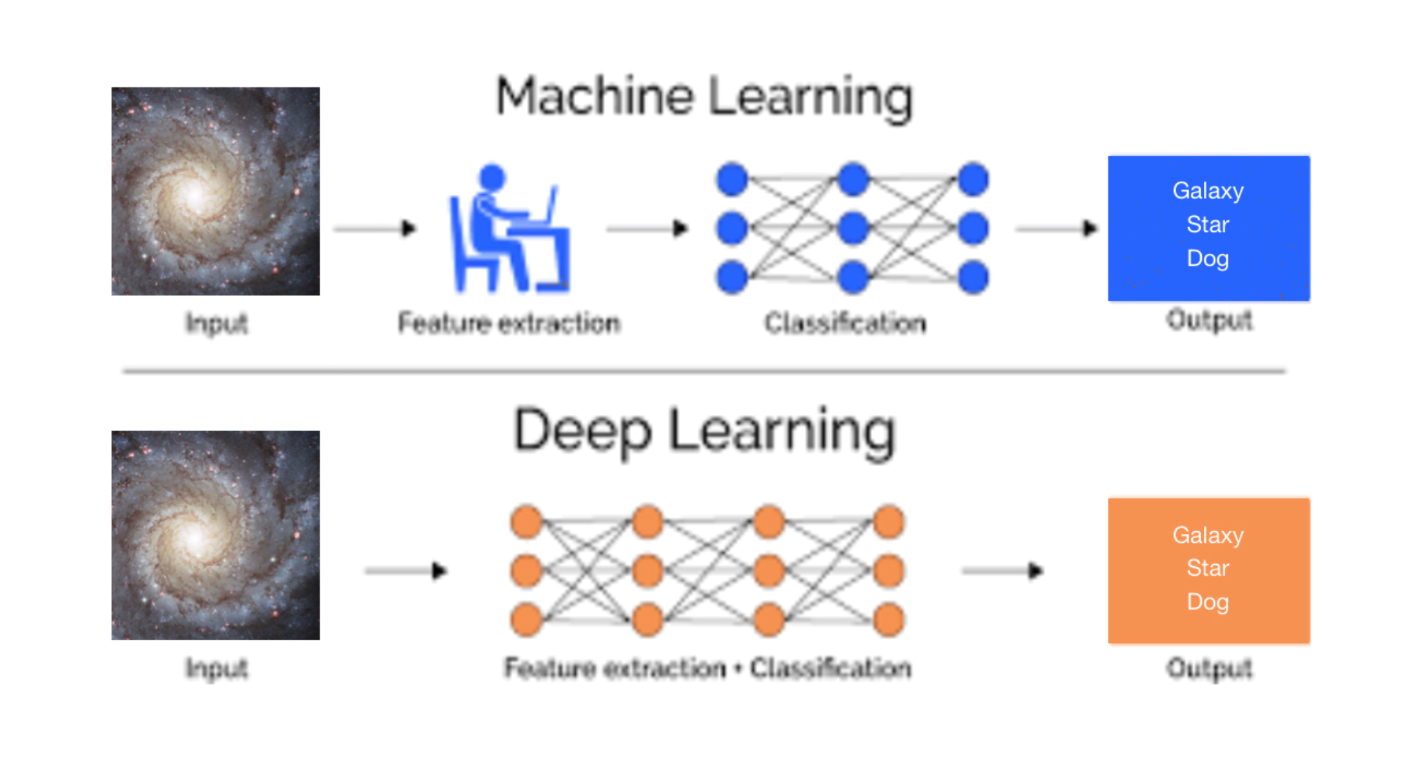}}

  \hspace*{.2in}
  \end{minipage}
  \vspace*{+3 mm}
 \caption{Contrasting `Shallow Learning'  (based on pre-processing feature extraction)  with `Deep Learning'. Graphics credit:  S. M. Lahav, based on Wikipedia}
 \label{fig2}
\end{center}
\end{figure}

Astrophysics, High Energy Physics and essentially all fields in Science, Technology and Medicine are undergoing an ‘industrial revolution’, with huge data sets and complex models. For example, the above mentioned galaxy surveys will catalogue billions of galaxies over the coming decade. Similarly, Particle Physics experiments, e.g. at CERN, generate gigantic data sets. Such data sets call for an entirely new approach for data analysis. Fortunately, powerful methods have been invented by statisticians and computer scientists, and they are widely applied across different fields. In particular,  ML techniques have  gained a lot of popularity in recent years.
Commonly used approaches are e.g. Artificial Neural Networks (ANN, see e.g. Figure 7), Boosted Decision Trees (BDT), and Support Vector Machines (SVM).  See Various methods implemented e.g. in packages such as Scikit- Learn\footnote{http://scikit-learn.org/stable/} and TensorFlow\footnote{https://www.tensorflow.org/}. A major breakthrough of recent years in ML is ‘Deep Learning’ (DL), see e.g. Le Cun et al. (2015) and Goodfellow et al (2016). 
Rather than pre-processing to preform feature extraction first. For example, some DL algorithms implement a cascade of multiple layers (`filters'), which allow the full data to ‘speak for themselves’ (see e.g. Figure 5). DL can be used for both supervised and unsupervised learning. A popular implementation of DL is Convolutional Neural Networks (CNN). DL is certainly a major new trend in Astrophysics  (Figure 3), with an increase from  23 papers with  DL in their titles in 2017,  to 105 in 2021. 
An example of a novel DL application  is the detection of strong gravitational lensing (Metcalf et al. 2019), where CNN wins over visual inspection and other ML algorithms.

While the performance of DL methods is impressive, it is still important to understand the DL algorithms, not to use them as just `black boxes’. This can be done by running the ML algorithms through toy models, as well using {\it explainability}  and {\it interpretability} tools (see below). 
We also argue that the traditional pre-processing of feature extraction in ML is not at all `shallow'. A lot of human knowledge of Physics and Astronomy has influenced the choice of such features.  
 
\section{Object classification}
\begin{figure}
\begin{center}
  \begin{minipage}{160mm}
  \centering
  \raisebox{-0.5\height}{\includegraphics[width=160mm, angle=0] {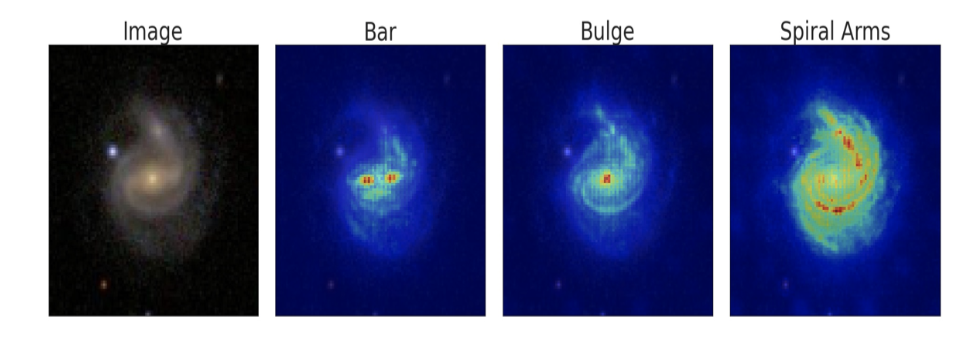}}
  \hspace*{.2in}
  \end{minipage}
  \vspace*{+3 mm}
  \caption{Example of explainable AI: The three saliency maps for the galaxy image on the left that indicate the sensitivity of each pixel to to the visually-classified features (bar, bulge and spiral arms), using an Explainable AI tool (Bhambra et al. 2022).}
   \label{fig2}
\end{center}
\end{figure}

Starting with a personal remark, 
I got interested in ML in the early 1990s, while  at the Institute of Astronomy in Cambridge. I discussed with colleagues the possibility that computers, rather than humans, would classify galaxies. A chance encounter with a book about ANN 
was followed by work with Michael Storrie-Lombardi and Laerte Sodre, and later  by the PhD work of Avi Naim (Storrie-Lombardi et al.1992; Lahav et al. 1995, Naim et al. 1995).
We defined an experiment of asking 6 classification ‘gurus’ to label 840 galaxy images (scanned by the Cambridge Automatic Plate Measuring  machine). An ANN was then trained to use a dozen or so
features to predict the de Vaucouleurs' T-system of 15 types. The ANN reproduced the classification to within 2 types.
This work was outside the main-stream at the time, but was later extended for classification of the Galaxy Zoo (e.g. Banerji et al. 2010, Walmsley et al. 2020).
It is an interesting question what features in the full image make the human decide on what type or another.
A step towards this is by Explainable AI tools, such as SmoothGrad. Bhambra et al. (2022)  have illustrated this by deriving saliency maps (see Figure 6), by calculating in each pixel the smoothed gradient of the  visually-classified T-type with respect to the pixel intensity (bypassing the internal architecture of the algorithm).

Another important area is the classification of galaxy spectra. Examples include Principal Component Analysis (e.g. Madgwick et al. 2003) and information theory  measures such as Entropy and Relative Entropy  (Slonim et al. 2000; Ferreras et al. 2022). Most methods show that the information content of a spectrum with thousands of bins can be compressed by a factor of 100 or so. It remains to be seen if advanced ML methods can unveil more information.

 Classification can also be done in the time domain. For example,  Lochner et al. (2016) have shown, using DES simulated light curves of Supernovae (SN) Type Ia how to extract features (using SALT2, parametric models and Wavelet) from SN light curves and to use five ML methods (Naïve Bayes, K Nearest-Neighbours, Support Vector Machine, ANN  and Boosted Decision Trees) to classify them. Using the area under the curve (AUC) they have found for example that Wavelet feature extraction with the BDTs algorithm, give an AUC of 0.98, where 1 represents perfect classification. This has evolved into the SNMachine software package, to be used by Rubin-LSST, expected to detect
 millions of supernovae in 10 years of observations. Another example is the  ML search for core collapse supernovae in gravitational wave data (Iess et al. 2023).

\section {Photometric redshifts}
\begin{figure}
\begin{center}di
  \begin{minipage}{140mm}
  \centering
  \raisebox{-0.5\height}{\includegraphics[width=70mm, angle=0] {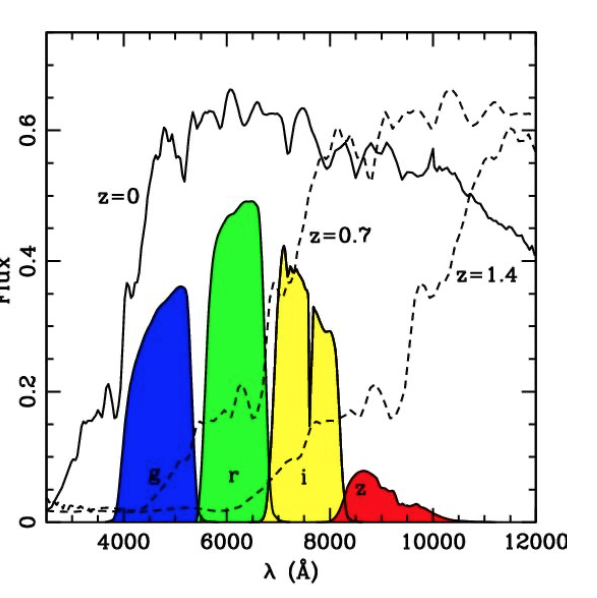}}
  \hspace*{.2in}
 \raisebox{-0.5\height}{\includegraphics[width=70mm]{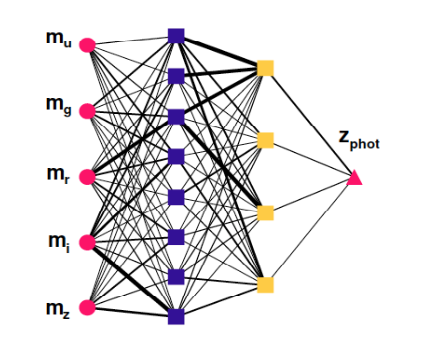}}
  \end{minipage}
  \vspace*{+3 mm}
  \caption{Top: A typical galaxy spectrum (flux against wavelength) at three  different redshifts, as samples by four filters g,r,i,z. Graphics credit:Wikipedia.
  Bottom: An Artificial Neural Network configuration with 5 magnitudes (one per filter) and the redshift as  output. Graphics credit: Sadeh et al. (2016).}
 \label{fig2}
\end{center}
\end{figure}

 Measuring the distance to galaxies is essential for cosmological studies such as galaxy clustering or weak gravitational lensing.
 Ideally, one would like the have a spectroscopic redshift  for each galaxy. In reality, getting spectra requires long exposure times,  so instead one can conduct multi-band photometry. Photometric redshifts (`photo-z') are not as precise as spectroscopic redshifts, but with many more galaxies with photometric redshifts than with spectroscopic redshifts  it is possible to get reliable cosmological results. Photometric surveys such as DES, KiDS, HSC, Rubin-LSST, Euclid and Roman-WFIRST heavily rely on the photometric redshifts.
It is probably fair to say that  the deduction of photo-z  is a pain for cosmologists (who would much prefer to use spectroscopic redshifts), but a joy for machine learners!  Indeed, photo-z  can be considered as an inverse problem,  going from flux measured through several filters to the redshift (see Figure 7) and its associated  probability distribution function. 
 There are over a dozen methods in the literature for deriving photometric redshifts. Broadly speaking they are divided into template and training methods, where a small spectroscopic redshift data set can be used to train the ML tool.  
In our group we  developed  two decades ago an Artificial Neural Network software package starting from ANNz 
(Collister and Lahav 2004) and then ANNz2 (Sadeh et al. 2016). We've also  attempted to improve the photo-z by adding galaxy structural parameters (Soo et al. 2018)
and by a Deep Learning approach of analysing the full galaxy image (Henghes et al. 2022, following Pasquet et al. 2019).
Furthermore, the photometric data can be used simultaneously to produce photo-z and stellar mass (Mucesh et al. 2021).
We've also explored the scalability of ML algorithms for photo-z with  the size of training data sets and other factors (Henghes et al. 2022).
For other ML approaches to the photo-z  problem, e.g. Self Organising Maps, see the review by Huertas-Company \& Lanusse  (2022).
Photo-z with ML remain an active field for many more years,  for Rubin-LSST, Euclid and other large photometric surveys.

\section {Simulation-based inference of galaxy masses,  dark matter maps and  cosmological parameters}

The advancements of both large simulations and ML algorithms have opened many new directions for research in Cosmology. Here are a few  examples:

{\bf (i)  The sum of masses of the Milky Way and Andromeda:} This was initially estimated by the classical `timing argument' (Kahn \& Waltjer 1959; Lynden-Bell 1981), assuming infall of two point-like objects towards each other. The combined mass is derived based on the observed present-epoch distance separation, the infall velocity and the age of  the universe. 
But in reality the two galaxies are part of the Local Group, hosting tens of galaxies, and  itself residing in the cosmic  web. 
To take into account this complexity of structure and dynamics one can compare the observed  two-body system with many  simulated galaxy pairs.
McLeod et al. (2017) have trained  Artificial Neural Networks  on 30,000 simulated galaxy halo pairs, and  Lemos et al. (2021) utilized Density Estimation Likelihood Free Inference with 2 million galaxy halo  pairs, extracted from LCDM simulations. 
The latter study derived for the combined mass  enclosed within density contrast of 200,  $M_{200} = 4.6 \times 10^{12\;\; +2.3}_{\;\;\;\;\;-1.8}$ (68\% CL) solar masses. The estimated value is in accord with the timing argument, but somewhat larger than the sum of individual masses derived by other method. The error bars are much more refined due to the advanced ML approaches employed.

\begin{figure}
\begin{center}
  \begin{minipage}{140mm}
  \centering
  \raisebox{-0.5\height}{\includegraphics[width=160mm]{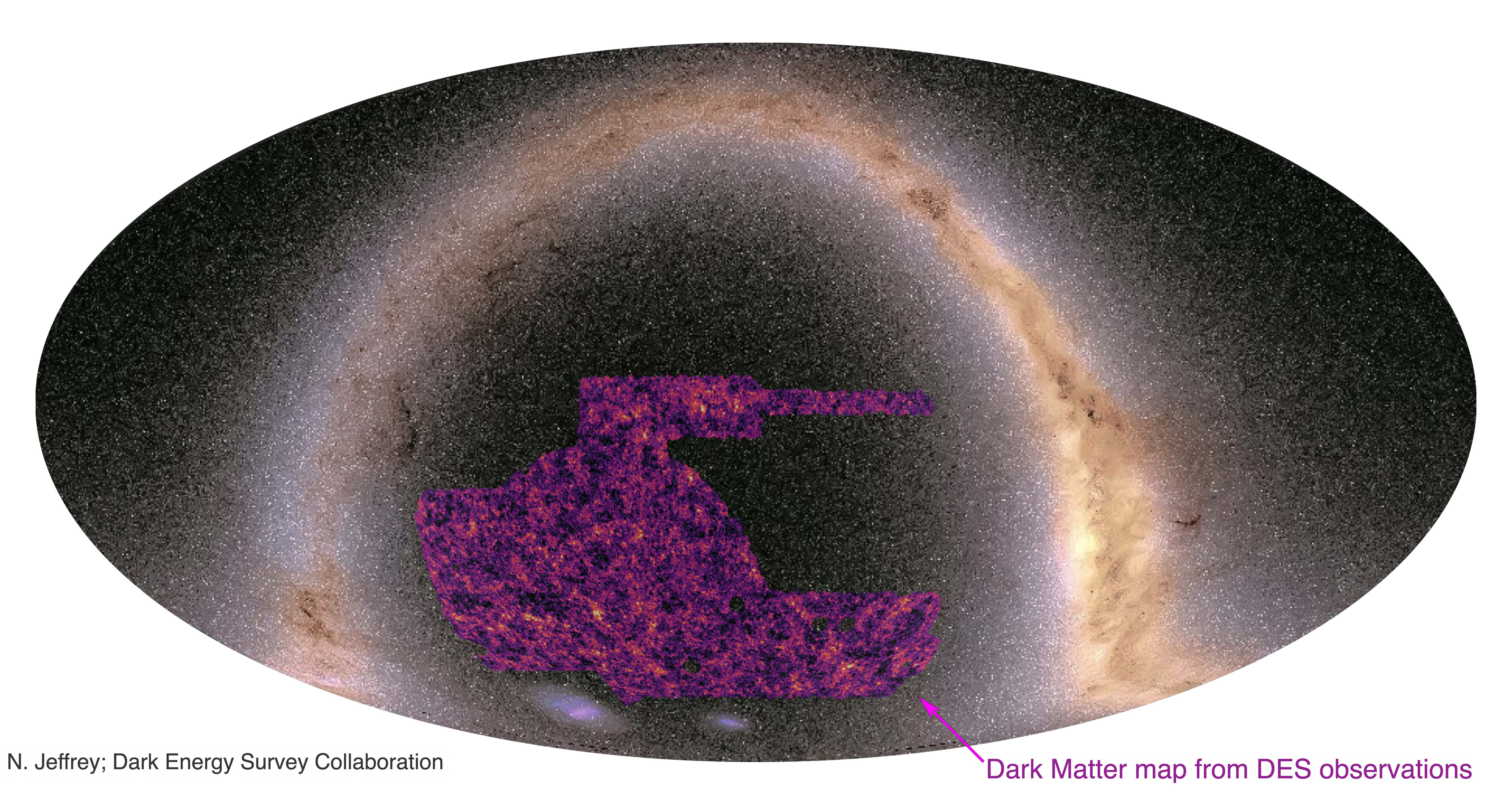}}
   \end{minipage}
  \vspace*{+3 mm}
  \caption{Mass map derived from DES Year 3 Weak Lensing, using 100 million galaxies with shape measurements over the 5000 survey area.   
  High and low density regions are shown in bright and dark colours, respectively.  The Milky Way image from Gaia is shown for the sky orientation.
  Credit: Nial Jeffrey \& DES collaboration (2021).}
   \label{DES_map}
\end{center}
\end{figure}

{\bf (ii) Reconstruction of dark matter maps from weak lensing:}
The inverse problem of deriving the projected mass distribution from the images of  distant galaxy images distorted (`cosmic shear') due to weak gravitational lensing  was addressed by the seminal algorithm of Kaiser \& Squires (1993). An example of a reconstructed mass map (using a Wiener filter)  is shown in Figure  8.  Several approaches to this reconstruction problem have been taken in recent  years, among them a Deep Learning approach technique (Jeffrey et al. 2021b). The algorithm learns the relationship between cosmic shear to 
projeted mass from about $10^5$ simulations. Then the observed shear (e.g, 100 million galaxy shape measurements from DES over 5000 sq deg of the sky) is presented to the algorithm to produce a dark matter map. These techniques can be applied further to Euclid and Rubin-LSST (each of the two covering about 20000 sq. deg., with expected 1 billion galaxies).
Such dark matter map are very useful for comparison with clusters and voids in galaxy maps (Hang et al., in preparation), and other multi-wavelength observations of the large scale structures.
More generally, maps are very useful for studying structure at the 'field level', beyond the commonly used 2-point statistics. The galaxy and dark matter maps can be used for N-point statistics (e.g. Gaudi et al. 2019), probability distribution functions (e.g. Clerkin et al. 2016) and graph statistics such as Minimum Spanning Tree (Naidoo et al., 2020). 
 On small scales, weak lensing maps can uncover the underlying dark matter distributions of halos and galaxy clusters. This can be used to study the effect of baryons and neutrinos on halo substructure, and particle interactions of dark matter on merging galaxy clusters. 
 Another promising direction is to use DL to train on maps to predict cosmological parameters 
(Kacprzak and Fluri 2022), Jeffrey et al. 2021a), and to even explain, via saliency maps (similar to Figure 6) the regions in the map that tell us most about certain cosmological and astrophysical parameters).
It is important  that the simulations are as realistic as possible and the systematics are modelled properly. 

{\bf (iii) ML for gravitational physics:}
Another growing area is in analyses of N-body and hydrodynamic simulations. For example, Lucie-Smith et al. (2018)  used ML to  study the relationship between the initial conditions in simulations and the final dark haloes.  ML can enhance large number of 'cheap' simulations, by training on a  small number of 'expensive' simulations, e.g. the Lagrangian Deep Learning 
of Dai \& Seljak (2021).
An even more ambitious goal is to let an ML algorithm to discover new physics, for example re-discovering the orbital mechanics  solar system (Lemos et al. 2022).  

\section{ Training the next generation of astronomers as data scientists}
The fast-moving AI technology requires a new type of training, for a dual career in Astrophysics (High Energy Physics or  other branches of Physics) as well as in Data Intensive Science (DIS).
As an example, the UK's STFC has initiated the new scheme of Doctoral Training Centre (CDT) in  DIS. 
Our  CDT-DIS at UCL\footnote{https://www.ucl.ac.uk/data-intensive-science-cdt/} trains about 10 CDT  PhD students per year.  
This is  a 4-yr PhD programme, including a 6-month placement in  one of 30 or so  private, public or third sector DIS organisations, e.g. from  Natural Language Processing for fashion retail  to ML for nuclear fusion. The PhD programme is now part of our UCL Centre for  Data Intensive Science and Industry (DISI), which also include a one-year MSc programme for about 70 students.
The Centre has strong links with international Astronomy and HEP projects and with other universities. With the support of the UK's Newton's Fund we also run a DIS  training programme with Jordan\footnote{https://www.ucl.ac.uk/astrophysics/research/ucl-jordan-dis-collaboration}.

\section{Outlook} 

AI is transforming the world. We have already reached the stage where we cannot analyse cosmological experiments without ML of some sort. ML certainly enhances productivity.
The relatively new development of Deep Learning has already proved  very useful for problems in Astronomy such as 
object classification, photometric redshifts, anomaly detection, enhanced simulations,  and inference of cosmological parameters. 
However, I argue that  the more traditional `shallow learning' (i.e. with pre-processing feature extraction) is actually quite deep, as it brings in human knowledge, while `deep learning'  might  be perceived as a black box, unless supplemented by explainability tools. 
The relationship of AI, laws of physics and human knowledge is interconnected, with cross-fertilisation.
As advocated by some,  AI  can  be viewed  as  `Augmented Intelligence',  rather than as  `Artificial Intelligence'.
In this context AI is enhancing human intelligence rather than replacing it (e.g.  Lochner \& Bessett 2021).

There are other challenges:  Can we explain/interpret ML algorithms without using them just as `black boxes'? How to minimise biases due to incomplete training sets?
Can we include prior Physics? Can we learn new Physics?
How to best train the next generation of astronomers also as data scientists?
This takes us to the next level, applying novel AI and ML methods  galaxy surveys (DES, DESI, Rubin-LSST, Euclid and others), with a huge discovery potential, from detecting subtle deviations in the standard LCDM model, measuring the neutrino mass and possibly discovering new populations of astronomical objects. 
It is probably fair to say that we are still missing  `killer applications'  of ML in cosmology, something like the recent success of ML for Protein Folding\footnote{
https://www.deepmind.com/research/highlighted-research/alphafold}.

Finally, I have asked the AI chatbot ChatGPT\footnote{chat.openai.com} the question posed by  this article's title: "Is Deep Machine Learning for Cosmology an evolution or a revolution?".
The chatbot has replied:

{\it ``Deep machine learning has the potential to be both an evolution and a revolution in the field of cosmology. On one hand, it represents an evolution in the sense that it builds upon and extends existing methods and approaches that have been used in cosmology for decades. For example, machine learning algorithms can be used to analyse and interpret large datasets in ways that are more efficient and accurate than traditional methods.

On the other hand, deep machine learning also has the potential to be a revolution in cosmology by enabling the discovery of new patterns and insights that were previously hidden or difficult to discern. For example, machine learning algorithms can be used to identify complex correlations and patterns in cosmological data that may not be easily identifiable through traditional methods of analysis. In this way, deep machine learning has the potential to fundamentally change the way that cosmologists approach and study the universe." }

\bigskip

Let's hope the ChatGPT's prediction in their last sentence is correct!

\bigskip
\bigskip

{\bf Acknowledgements}
I thank my PhD students, Post-docs  and collaborators over the years for their contribution to the  work presented here.
I acknowledge an Advanced ERC and STFC consolidated and CDT-DIS  grants for their support of this research, and a Visiting Fellowship at 
All Souls College Oxford.

\bigskip
\bigskip

{\bf References}


Banerji, M., et al., 2010, MNRAS, 406, 342

Baron, D., et al., 2019, arXiv:1904.07248

Bhambra, P., et al, 2022, MNRAS, 511, 5032

Bishop, C., 2006, {\it Pattern recognition and Machine Learning}, Springer

Calder, L. \& Lahav, O., 2008, RAS Astronomy \& Geophysics,  49, 1.13

Carleo, G., et al., 2019, Rev Mod Phys, 91, 045002

Chang, C., et al., 2016, MNRAS, 459, 3203

Clerkin, L., et al., 2017, MNRAS, 466, 1444

Collister, A. \& Lahav, O., 2004, PASP, 116, 345

Dai, B. \& Seljak, U., 2021, PNAS, 118 (16) e2020324118

Dark Energy Survey  collaboration, 2016, MNRAS, 460, 1270

Dark Energy Survey collaboration, 2022, Phys Rev D, 105, 023520 


Elgaroy, O,. et al., 2002, Phys  Rev Lett, 89, 061301

Ferreras, I., et al.,  2022, RASTI (in press), arXiv:2208.05489

Goodfellow, I., et al. ., 2016, {\it Deep Learning}, MIT Press


Gualdi, D., et al., 2019, MNRAS, 484, 3713

Henghes, B., et al., 2021, MNRAS, 505, 4847 

Henghes, B., et al., 2022, MNRAS, 512, 1696

Huertas-Company, M. \& Lanusse, F.,  2022,  arXiv: 2210.01813

Iess, A., et al., 2022, A\&A, 669, A42

Jeffrey, N., et al.,  2021a, MNRAS, 501, 954

Jeffrey, N., et al.,  2021b, MNRAS, 505, 4626

Kacprzak, T. \& Fluri, J.,  2022, arXiv:2203.096116

Kahn, F.D., \& Woltjer, L., 1959, ApJ, 130, 705 

Kaiser, N. \& Squires G., 1993, ApJ, 404, 441
 
Lahav, O., et al., 1995, Science, 267, 859

Lahav, O. \& Liddle, A., 2021, Reviews of Particle Physics, arXiv:2201.08666

Lahav, O. \& Silk, J., 2021, Nature Astronomy, 5, 855

Lahav, O., et al. (eds.), 2020, {\it The Dark Energy Survey:
The Story of a Cosmological Experiment}, World Scientific

Le Cun, Y., et al. , 2015, Nature, 521, 43

Lemos, P., et al., 2021,  Phys. Rev. D,  103, 023009

Lemos, P., et al., 2022, arXiv:2202.02306

LIGO collaboration, 2017a, ApJ, 848, L12

LIGO collaboration, 2017b, Nature, 551, 85

Lochner, M., et al. , 2016, ApJS, 225, 31

Lochner, M.  \& Bassett, B.A., 2021, Astronomy and Computing, 36, 100481

Lucie-Smith, L., et al., 2018, MNRAS, 479, 3405

Lynden-Bell, D., 1981, Observatory, 101, 111


Madgwick, D.S.,  et al., 2003, MNRAS, 343, 871

McLeod, M., et al., 2017, JCAP, 12 03

Metcalf, R.B.,  et al., 2019, A\&A, 625, A119 

Mucesh, S.,  et al., 2021, MNRAS, 502, 2770

Naidoo, K., et al., 2020, MNARS, 491, 1709

Naim, A.,  et al. 1995, MNRAS, 275, 567

Pasquet,  J., et al., 2019, A \& A,  621, A26

Perlmutter, S., et al. 1999, ApJ, 517, 565

Planck collaboration, 2018, arXiv:1807.06209

Riess , A.. et al., 1998, AJ, 116, 1009

Riess,  A., et al., 2021, ApJ, 853, 126

Sadeh, I., et al..  2016, PASP, 128, 4502 

Shah, P., et al., 2021, A\&A  Review, 29, 9 

Slonim, N., et al. 2001, MNRAS, 323, 270

Soo, J., et al., 2018, MNRAS, 475, 3613


Storrie-Lombardi,  M..C., et al., 1992, MNRAS, 259, 8p

Walmsley , M.,  et al., 2020, MNRAS, 491, 1554









\end{document}